\documentstyle[psfig]{l-aa}  

\begin{document}

\thesaurus{03(08.09.2 GRS 1915+105; 11.01.2; 11.10.1; 11.14.1; 11.09.1
NGC 4258; 02.01.2)}
\title{The jet/disk symbiosis III.}
\subtitle{What the radio cores in GRS 1915+105, NGC 4258, M81, and Sgr A
tell us about accreting black holes}
\author{Heino Falcke \and Peter L. Biermann}
\institute{
Max-Planck Institut f\"ur Radioastronomie, Auf dem H\"ugel 69, D-53121
Bonn, Germany
}
\offprints{hfalcke@mpifr-bonn.mpg.de}
\date{Astronomy \& Astrophysics, in press}

\markboth{Falcke \& Biermann: Jet-Disk Symbiosis III. - Application to GRS~1915+105, NGC~4258}{Falcke \& Biermann: Jet-Disk Symbiosis III. - Application to GRS~1915+105, NGC~4258}
\maketitle
\begin{abstract}
We have derived simplified equations for a freely expanding, pressure
driven jet model as a function of jet power and applied it
successfully to the radio cores in the black hole candidates
GRS~1915+105, NGC~4258, and M81 which are observationally well defined
systems, and to Sgr A*. By using equipartition assumptions, the model
has virtually no free parameters and can explain all sources by just
scaling the jet power.  In GRS~1915+105 it also naturally explains the
jet velocity and the radio time delay. The jet powers we derive for
the radio cores of the first three sources are comparable to their
accretion disk luminosities, providing further evidence for the
existence of symbiotic jet/disk systems and a common engine mechanism
also in low-luminosity AGN and stellar mass black holes. With the
exception of Sgr~A* an advection dominated accretion flow (ADAF) does
not seem to be necessary to explain any of the radio cores which span
a large range in luminosity and size, as well as in black hole masses
and accretion rate---from Eddington to extreme sub-Eddington. We
suggest, however, that the jet model can be used to derive minimum
accretion rates and thus find that Sgr A* seems to be truly
radiatively deficient---even in a starved black hole model---and that
a combination of jet and ADAF model may be one possible solution.
\end{abstract}
\section{Introduction}

Early on in the discussion about the existence of black holes
Lynden-Bell \& Rees (1971) suggested that they would be accompanied by
compact radio nuclei, detectable by Very Long Baseline Interferometry
(VLBI), and predicted such a source for the Galactic Center. Indeed,
this source (Sgr A*) was then discovered by Balick \& Brown (1974) and
it became clear in later years that compact radio cores are indeed
good evidence for the existence of active galactic nuclei (AGN) most
likely powered by black holes. For luminous radio galaxies and
radio-loud quasars the basic nature of these compact radio nuclei has
been clarified in the meantime through extensive and detailed VLBI
observations (see Zensus 1997 for a review) as being the inner regions
of relativistic jets emanating from the nucleus.

Despite this progress, a number of important questions remain when
looking back at the initial discussion. First of all, it is unclear
whether there indeed is a direct link between compact radio cores and
AGN, i.e.~whether compact radio cores and jets are just an accidental
by-product of black hole activity or a necessary
consequence. Secondly, for the lesser studied, low-luminosity AGN the
jet nature of compact radio nuclei has not yet been established beyond
any doubt, leaving the question open whether in fact a compact radio
core in a low-luminosity AGN (LLAGN) is the same as in a
high-luminosity AGN, i.e. a quasar.

The latter was exactly the claim we made in an earlier paper (Falcke
\& Biermann 1995) where we proposed that accretion disks and jets form
symbiotic systems and proposed a scaling law which connects high-power
and lower-power accretion disks and their associated radio jets
(cores). The scaling law was based on the assumption of an
equipartition between the energy released and radiated away through
dissipation processes in the accretion disk and the power put into the
formation of magnetically driven radio jets.

The question whether this scaling law holds all the way down to
low-luminosity AGN, as claimed in Falcke \& Biermann (1996), also has
some very interesting implication for the current discussion of
accretion flows.  Since the early papers on the observational
appearance of black holes (e.g.~Shakura \& Sunyaev 1973) it was
assumed that luminous, thermal emission at optical, ultra-violet (UV),
or X-ray wavelengths was the primary sign for the presence of an
accreting black hole. It was argued that any matter falling onto the
black hole would likely form an accretion disk, if there was any
residual angular momentum, and hence would need to dissipate its
potential energy into heat by viscous processes allowing it to
transport angular momentum outwards while matter is falling inwards
($\alpha$-disk). This idea was used successfully to explain the ``big
blue bump'' in quasars (e.g. Sun \& Malkan 1998).

However, the view that the $\alpha$-disk can be extended to much lower
powers has been challenged (Narayan \& Yi 1994, 1995a\&b) and it was
argued that accretion disks will turn into advection dominated
accretion flows (ADAFs) if the accretion rate onto the black hole is
sufficiently sub-Eddington.  Narayan et al.~(1995 \& 1998; see also
Rees 1982 and Melia 1994) applied this to the Galactic Center, trying
to explain the compact radio source Sgr A* and its faintness at other
wavelengths, and Lasota et al.~(1996) used the ADAF model to explain
the broad-band spectrum of the nearby LLAGN and LINER galaxy NGC~4258
which is famous for its megamaser emission from a molecular disk
(Miyoshi et al.~1995). An integral part of these ADAF models is the
prediction of very compact radio emission associated with the
innermost part of the accretion flow, providing an alternative
explanation to the jet model for compact radio nuclei in LLAGN. While
initially the predicted, highly inverted radio spectra of the ADAF
model, did not fit the observed characteristics of these radio cores
very well, Mahadevan (1998) presented a more recent version of this
model that was at least able to account for the correct radio spectrum
of Sgr~A*\footnote{Interestingly, the proposed process (pion decay in
pp-collisions) had also been proposed and used within the jet model
for the Galactic Center (Falcke 1996a)}. Still, Di~Matteo et
al.~(1998) found a number of serious constraints for ADAF models---at
least for compact radio nuclei in elliptical galaxies.

Hence, the question now is whether indeed the radio emission from
compact nuclei in sub-Eddington accretion systems can be used as an
argument for the existence and necessity of ADAFs, or whether they are
equally well, or even better explained, in a scaled down AGN jet
model.  The latter will be tested in this paper: firstly, we will use
the jet/disk-symbiosis model of Falcke \& Biermann (1995) in its most
recent version (Falcke 1996b) and present simplified approximate
solutions that can be applied easily. Secondly, we will apply those
solutions to some specific sources which are of particular interest in
this discussion and are observationally well constrained in their
parameters. Finally we will discuss our results within the context of
the jet/disk-symbiosis model and their implications for ADAFs.

\section{The jet/disk-symbiosis model}
\subsection{Basics}

The model by Falcke \& Biermann (1995) was derived from a simple
Blandford \& K\"onigl (1979) model which calculates the synchrotron
emission of a relativistic, conical jet as a function of jet power
$Q_{\rm jet}$ (here: of two cones) parametrized by the accretion disk
luminosity $L_{\rm disk}$, such that $Q_{\rm jet}=q_{\rm j/l} L_{\rm
disk}$. Here we will make use of this parametrization only to
recapitulate our earlier results and later express the simplified
equations in terms of the jet power alone. In the end we will
therefore be able derive the parameter $q_{\rm j/l}$ without making
any ab-initio assumptions about its value.

The maximally efficient model assumes equipartition between magnetic
field and particles and between internal and kinetic energy
($\beta_{\rm s}=\sqrt{(\Gamma-1)/(\Gamma+1)}\sim0.4$). In Falcke
(1996b) we added a self-consistent description of the velocity field:
the jet was considered to be a fully relativistic electron/proton
plasma with a turbulent magnetic field treated as a photon gas leaving
a nozzle and freely expanding into the vacuum (this is in contrast to
large scale jet models which assume pressure equilibrium and confining
cocoons). The longitudinal pressure gradient then leads to a modest, yet
significant acceleration along the z-axis of the jet in the asymptotic
(i.e. observable) regime given by the equation for the jet proper
velocity $\gamma_{\rm j}\beta_{\rm j}$

\begin{equation}\label{euler2}\label{v}
\left({\left({\Gamma+\xi\over\Gamma-1}\right)(\gamma_{\rm j}\beta_{\rm
j})^2-\Gamma\over\gamma_{\rm j}\beta_{\rm j}}\right)
{\partial\gamma_{\rm j}\beta_{\rm j}\over\partial z}
={2\over z}
\end{equation}
with $\xi=\left(\gamma_{\rm j}\beta_{\rm
j}/(\Gamma(\Gamma-1)/(\Gamma+1))\right)^{1-\Gamma}$ and an adiabatic
index $\Gamma=4/3$ (Falcke 1996b).

This model assumes that there is no additional acceleration beyond the
nozzle and therefore just gives a lower limit to the terminal jet
speed of order $\gamma_{\rm jet}=2-3$ which seems to be just enough
for some low-power nuclei. This will of course fail for powerful
quasar jets since they exhibit significantly faster motion. We also
point out that this does not make any assumption on the process of jet
formation itself which is here treated as a ``black box'' of linear
dimension $Z_{\rm nozz}$ so that $z=Z/Z_{\rm nozz}$, where $Z$ is the
distance from the black hole.

The radio spectrum of such a pressure driven jet (which is no longer
perfectly conical) can then be calculated assuming energy conservation
along the jet when assuming a certain energy distribution of the
relativistic electrons as described in more detail in Falcke
(1996b). For simplicity we assume that a fraction $x_{\rm e}$ of the
electrons gets accelerated up to an initial characteristic energy of
$\gamma_{\rm e}m_{\rm e}c^2$ (or possibly produced near this energy by
$\pi^{\pm}$-decay, e.g. Biermann et al.~1995). While the existence of
quasi-monoenergetic electron distributions has been claimed for LLAGN
(Duschl \& Lesch 1994; Reuter et al.~1996), this choice here was made
mainly to account for the possibility of a low-energy cut-off/break in
the electron distribution (Celotti \& Fabian 1993; Falcke \& Biermann
1995). Since in almost all cases $\gamma_{\rm e}$ will have to be such
that the characteristic frequency $\nu_{\rm c}$ is close to the
self-absorption frequency $\nu_{\rm ssa}$, all the results
obtained here will also be roughly valid for a more realistic
power-law distribution with low-energy break at $\gamma_{\rm e}$.  On
the other hand it can also not be excluded that indeed electrons in radio
cores start out with a very narrow energy distribution when they are
injected and only along the way get accelerated into a power-law
distribution.

\subsection{Simplified equations} 

Using all the basic premises above, one arrives at a set of equations
for the expected radio spectrum and typical size scale of a radio core
for a given accretion disk luminosity (or jet power), inclination
angle and electron energy, given as Eqs. (8-12) in Falcke (1996b) for
the case of M81*, which we reproduce here (using the parameters
exactly as specified in that paper). The characteristic scale $z_{\rm
c}$ of the radio core is
\begin{eqnarray}
z_{\rm c}&=(z_{{\rm c,0}}/\sin i)(\mu_{\rm p/e}x_{\rm e})^{-2\xi} \left(q_{0.5}L_{41.5}\right)^{\xi/2}z_{13.7}^{1-\xi}\nu_{10.3}^{-\xi}
\end{eqnarray}
where the parameters are $z_{13.7}=z_0/3$~AU, $\nu_{10.3}=\nu/22$~GHz,
$\xi=(0.99, 0.95, 0.9, 0.89, 0.88)$ and $z_{{\rm
c,0}}=$(500, 1200,1100, 900, 700) AU for $i=(5^\circ, 20^\circ, 40^\circ,
60^\circ, 80^\circ)$, while the total 
spectrum of the jet is given by

\begin{eqnarray}\label{oldflux}
S_{\nu}&=745\,{\rm mJy}(q_{0.5}L_{41.5})^{1.46}\mu_{\rm
p/e}^{.17}x_{\rm e}^{1.17}z_{13.7}^{0.08}\nu_{10.3}^{0.08}\nonumber\\
&-337\,{\rm mJy}(q_{0.5}L_{41.5})^{1.54}\mu_{\rm p/e}^{.92}x_{\rm
e}^{2.1}z_{13.7}^{0.08}\nu_{10.3}^{0.08},\\
S_{\nu}&=247\,{\rm mJy}(q_{0.5}L_{41.5})^{1.42}\mu_{\rm
p/e}^{.33}x_{\rm e}^{1.33}z_{13.7}^{0.16}\nu_{10.3}^{0.16}\nonumber\\
&-143\,{\rm mJy}(q_{0.5}L_{41.5})^{1.48}\mu_{\rm p/e}^{.85}x_{\rm
e}^{1.95}z_{13.7}^{0.15}\nu_{10.3}^{0.15},\\
S_{\nu}&=106\,{\rm mJy}(q_{0.5}L_{41.5})^{1.40}\mu_{\rm
p/e}^{.40}x_{\rm e}^{1.40}z_{13.7}^{0.20}\nu_{10.3}^{0.20}\nonumber\\
&-71\,{\rm mJy}(q_{0.5}L_{41.5})^{1.45}\mu_{\rm p/e}^{.81}x_{\rm
e}^{1.89}z_{13.7}^{0.19}\nu_{10.3}^{0.19},
\end{eqnarray}
for $D=3.25$ Mpc, and $i=(20^\circ,40^\circ,60^\circ$)
respectively.

Unfortunately, these equations are not easy to handle since the
power-law indices are a function of the inclination angle. We
therefore derive here an approximate formula for the spectra and sizes
predicted by such a model.

We note here once more that Eq.~\ref{oldflux} has two inherent
constraints. First of all, by definition $\mu_{\rm p/e}$ cannot be
smaller than unity since it is defined as the ratio between the energy
densities in protons and electrons plus one. For the sake of
simplicity only, we will now ignore the relativistic proton content and
set $\mu_{\rm p/e}=1$, so that we can substitute the relativistic
electron fraction $x_{\rm e}$ with
\begin{equation}\label{xe}
x_{\rm e }=m_{\rm p}/\left(4\Gamma m_{\rm e}\gamma_{\rm e}\right)=344/\gamma_{\rm e}.
\end{equation}
Secondly, one cannot increase $\gamma_{\rm e}$ indefinitely since at
some point the flux would become negative, i.e.~the jet would become
completely self-absorbed and the simplifications would break down.
Moreover, the equations are difficult to handle because of this sum,
since for large changes in $Q_{\rm jet}$ the sum also would become
negative.  Hence,, we formally introduce an arbitrary scaling relation
\begin{equation}\label{gammae}
\gamma_{\rm e}=\gamma_{\rm e,0}\left({Q_{\rm jet}\over
10^{39}\mbox{erg/sec}}\right)^{0.09}
\end{equation}
which allows us to simplify the equations further. The physical
meaning is that electrons are pushed to somewhat higher energy
with increasing $Q_{\rm jet}$ to keep them in the optically thin
part. A mechanism which indeed could lead to such an effect is the
`synchrotron boiler' (Ghisellini et al.~1988) that describes the
evolution of low-energy electrons in a self-absorbed system, but it is
not clear whether this formally introduced relation here has any
significance in the real world and therefore we will ignore it in the
discussion of our results.

To further simplify the equations we have rounded the exponents
typically to the 2nd digit after the decimal and factorized the
equations. Even for the most strongly varying parameters, like $L_{\rm
disk}$ which can vary over 6 orders of magnitude, the resulting error
will be only some ten percent. Moreover, the exponents in the equations,
which are a function of the inclination angle $i$ of the jet, were
fitted by 2nd and 3rd order polynomials in $i$ to an accuracy of much
better than a few percent over a large range of angles.

All these simplifications lead to the following expressions for the
observed flux density and angular size of a radio core observed at a
frequency $\nu$ as a function of jet power. For a source at a distance
$D$, with black hole mass $M_\bullet$, size of nozzle region $Z_{\rm
nozz}$ (in $R_{\rm g}=G M_\bullet/c^2$), jet power $Q_{\rm jet}$,
inclination angle $i$, and characteristic electron Lorentz factor
$\gamma_{\rm e}$ (see Eq.~\ref{gammae}) the observed flux density
spectrum is given as

\begin{eqnarray}\label{flux}
S_{\nu}&&=10^{2.06\cdot\xi_0}\;{\mbox mJy}\;\left({Q_{\rm jet}\over10^{39} \mbox{erg/sec}}\right)^{1.27\cdot\xi_1}
\nonumber\\&&\cdot
\left({D\over10{\rm kpc}}\right)^{-2}
\left({\nu\over8.5 {\rm GHz}}\right)^{0.20\cdot\xi_2}
\left({M_\bullet\over33 M_\odot}{Z_{\rm nozz}\over10 R_{\rm g}}\right)^{0.20\cdot\xi_2}
\nonumber\\&&\cdot
\left(3.9\cdot\xi_3\left(\gamma_{\rm e,0}\over200\right)^{-1.4\cdot\xi_4}
-2.9\cdot\xi_5\left(\gamma_{\rm e,0}\over200\right)^{-1.89\cdot\xi_6}\right),
\end{eqnarray}
with the correction factors $\xi_{0-6}$ depending on the
inclination angle $i$ (in radians):

\begin{eqnarray}
\xi_0&=&2.38 - 1.90\,i + 0.520\,{i^2}\\
\xi_1&=&1.12 - 0.19\,i + 0.067\,{i^2} \\
\xi_2&=&-0.155 + 1.79\,i - 0.634\,{i^2} \\
\xi_3&=&0.33 + 0.60\,i + 0.045\,{i^2} \\
\xi_4&=&0.68 + 0.50\,i - 0.177\,{i^2} \\
\xi_5&=&0.09 + 0.80\,i + 0.103\,{i^2}\\
\xi_6&=&1.19 - 0.29\,i + 0.101\,{i^2}.
\end{eqnarray}
Likewise, the characteristic angular size scale of the emission region
is given by

\begin{eqnarray}\label{size}
\Phi_{\rm jet}&=&1.36\cdot\chi_0\,\mbox{mas}\,\sin{i}
\nonumber\\&\cdot&
\left(\gamma_{\rm e,0}\over200\right)^{1.77\cdot\chi_1}
\left({D\over10{\rm kpc}}\right)^{-1}
\left({\nu\over8.5 {\rm GHz}}\right)^{-0.89\cdot\chi_1}
\nonumber\\&\cdot&
\left({Q_{\rm jet}\over10^{39} \mbox{erg/sec}}\right)^{0.60\cdot\chi_1}
\left({M_\bullet\over33 M_\odot}{Z_{\rm nozz}\over10 R_{\rm g}}\right)^{0.11\cdot\chi_2},
\end{eqnarray}
with the correction factors

\begin{eqnarray}
\chi_0&=& 4.01 - 5.65\,i + 3.40\,{i^2} - 0.76\,{i^3}\\
\chi_1&=& 1.16 - 0.34\,i + 0.24\,{i^2} - 0.059\,{i^3}\\
\chi_2&=& -0.238 + 2.63\,i - 1.85\,{i^2} + 0.459\,{i^3},
\end{eqnarray}
where again the inclination angle $i$ is in radians. We point out that
in this model the characteristic size scale of the core region is
actually equivalent to the offset of the radio core center from the
dynamical center. This does not exclude the existence of emission in
components further down the jet, which might be caused by shocks or
other processes.

\begin{figure*}[t]
\centerline{\mbox{\psfig{figure=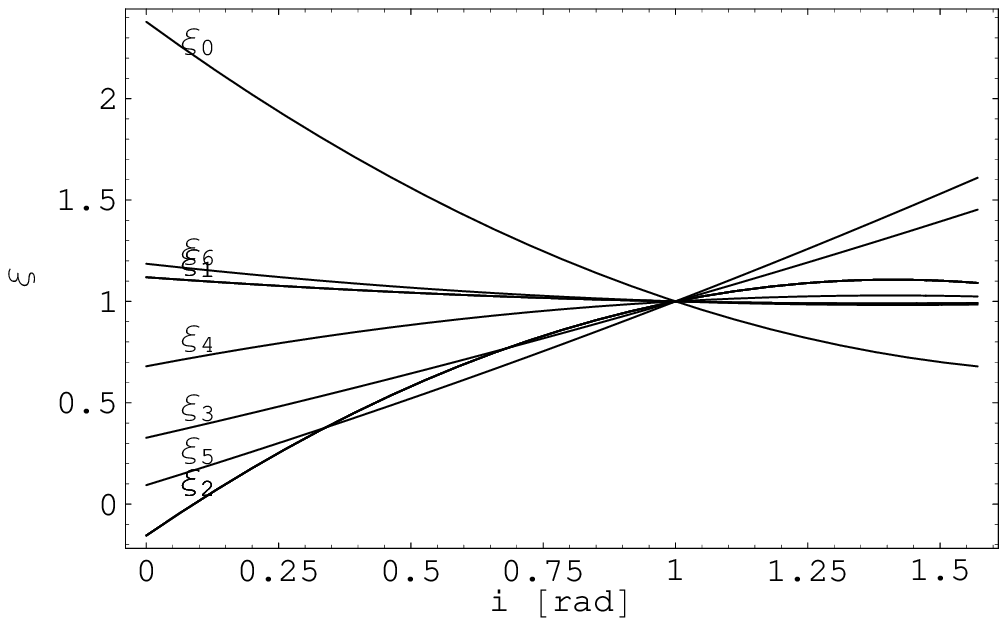,width=8.2cm,bbllx=3.25cm,bburx=13.5cm,bblly=20.2cm,bbury=27cm}\hfill\psfig{figure=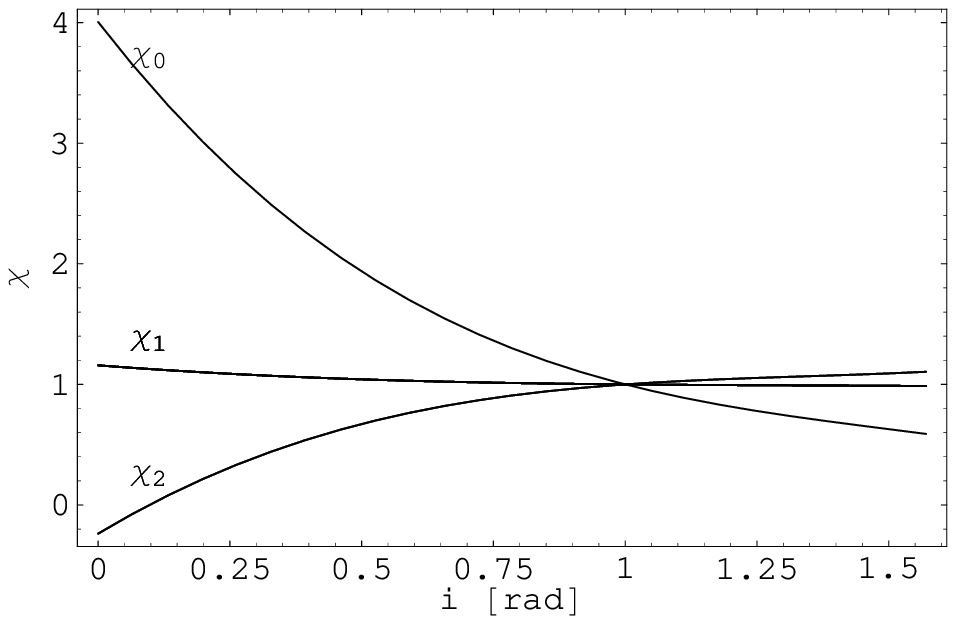,width=8.2cm,bbllx=3.25cm,bburx=13.5cm,bblly=20.2cm,bbury=27cm}}}
\caption[]{Correction factors $\xi$ and $\chi$ for the exponents and
fore-factors in Eqs.~\ref{flux} \& \ref{size} as a function of
inclination angle $i$ in radians, where $i=0$ corresponds to face-on
orientation. Note, however, that for $i\la10^\circ$ most
approximations fail.}
\end{figure*}

The equations are scaled to the typical values for Galactic jet
sources like GRS1915+105 and the correction factors are normalized to
an inclination angle of 1 rad ($\sim57^\circ$). We note that the
approximations fail at small inclination angles where the accretion
disk is seen face on and the jet points towards the observer
(i.e. $i\la10^{\circ}$). The benefits of these equations now are that
they can be used to quickly compare observed radio core properties
with the model predictions, especially since we have reduced the
number of free parameters to the absolute minimum.

\section{Application to individual sources}
Application of these simplified equations to real radio cores is
straight forward, the basic input parameters being the jet power, the
characteristic electron energy, the inclination to the line of sight,
the observed frequency, the distance, the black hole mass, and the
relative size of the nozzle region. The latter two enter only weakly
and hence need to be known only to an order of magnitude.

A few systems are so well studied that most of these parameters
(especially $i$, $D$, \& $M_\bullet$) can be fixed with some
confidence and where size and flux of their cores at a certain
frequency are well known through VLBI observations. Even though this
may be a subjective criterion, we believe that the radio cores in M81,
NGC~4258, and GRS~1915+105 are, for various reasons, perhaps the best
studied and best constrained examples of low-power radio
cores. Following the convention in Falcke (1996b) and Melia (1992) and
in analogy to Sgr A*, we will identify the radio cores in these
sources by adding an asterisks to their host galaxy or source name to
clearly distinguish them from their hosts.

We have listed the sources and their parameters in Table 1. The
observed quantities we have used as input parameters for the model are
given in Columns 2-7. Since in all cases, except Sgr~A*, we have only
two unknowns left (jet power and characteristic electron energy) to
describe the two observed quantities of the radio cores (flux and
size) we were able to solve the model equations for each source
completely and determine $Q_{\rm jet}$ and $\gamma_{\rm e}$ directly
from the observations. These results and the predicted spectral
indices for the radio spectrum are given in the three right-most
columns of Table 1. For comparison with the jet power, we also listed
the accretion disk luminosity of each system in Column 8. In the
following we will briefly discuss the data and the modelling of each
source.

\subsection{NGC~4258}
The VLBI observations of megamaser emission has led to
the detection of a molecular disk in NGC~4258 (Miyoshi et
al.~1995) which can be used to determine the inclination angle
$i=82^{\circ}$ of the system, the black hole mass
$M_\bullet=3.5\cdot10^7M_\odot$, and the distance $D=7.3\pm0.3$ Mpc
(Herrnstein et al.~1997a) almost directly from the observations. The
variable central VLA radio core (Turner \& Ho 1994), here called
NGC~4258*, has a flux of roughly 3 mJy and was interpreted by Lasota
et al.~(1996) as emission from an ADAF while Falcke (1997) suggested a
scaled down AGN jet origin. The latter picture was confirmed by
Herrnstein et al.~(1997b\&1998) who discovered a nuclear jet in
NGC~4258 offset by 0.35 to 0.46 mas from the dynamical
center. Herrnstein et al.~(1996) suggested that this offset could be
interpreted within the framework of the Blandford \& K\"onigl (1979)
model as being due to self-absorption in the inner jet cone. The
search for radio emission directly at the dynamical center remained
unsuccessful (Herrnstein et al.~1998) and required a revision of the
Lasota et al.~(1996) ADAF model (Gammie et al.~1998).

For our purposes NGC~4258* is an ideal system because all crucial
parameters, especially the inclination angle, seem to be fixed.  Using
an average radio flux of 3 mJy at 22 GHz and the offset of the core
from the dynamical center as the characteristic size scale of the
system we find a jet power of $10^{41.7}$ for the nuclear jet, a
characteristic electron Lorentz factor of $\sim630$, and predict an
average spectral index $\alpha=0.22$ ($S_\nu\propto\nu^\alpha$). The
jet-power of the nuclear jet is consistent with the large scale
emission-line jet in NGC 4258, since its kinetic power is also of the
order $10^{42}$ erg/sec --- as derived from the mass ($2\cdot10^6
M_\odot$) and velocity ($\sim2000$km/sec) of the emission-line gas
(Cecil et al.~1995). Moreover, this is also in line with the estimated
nuclear accretion disk luminosity of $\sim10^{42}$erg/sec (St\"uwe et
al.~1992; Wilkes et al.~1995; see also discussions in Herrn\-stein et
al.~1997 and Gammie et al.~1998). Hence, all the activity in NGC~4258
can be described in a consistent way by a low-luminosity
jet/disk-system and an accretion rate of the order $10^{-4}
M_{\sun}/$yr. One caveat exists, however, because the interpretation
of the offset of the core from the dynamical center as the
characteristic scale of the model (and not the self-absorption size
which is smaller in this model) actually implies that also the core
size is of similar order. If it were smaller, e.g. 0.1 mas, this would
reduce the, compared to other sources, relatively high value for
$\gamma_{\rm e}$ to around 200 without significantly reducing the
required jet power.  A difference between offset and actual core size
would occur if the jet were collimated in the inner region more than
assumed in our model (i.e.~were narrower than the Mach cone).

\subsection{GRS~1915+105}

Mirabel \& Rodriguez (1994) discovered a compact radio jet in
GRS~1915+105 with apparent superluminal motions, for which they were
able to determine the jet speed ($0.92c$) and the inclination angle
($\sim70^\circ$) of the system. Moreover, in recent papers Fender et
al. (1997), Pooley \& Fender (1997), Mirabel et al.~(1998) and
Eikenberry et al.~(1998) found an intriguing correlation between radio
outbursts and X-ray flares and hence a symbiotic jet/disk-system as
proposed in Falcke \& Biermann (1995\&1996) seems to be a good
description for GRS~1915+105.  The parameters $L_{\rm
disk}\sim10^{39}$erg/sec and $M_{\bullet}\sim33 M_{\sun}$ are
discussed in the literature (e.g.~Mirabel et al.~1997; Morgan et
al.~1997) for this source, but we point out that the mass
determination is extremely uncertain, yet is also not really critical
for the modelling.

Dhawan et al.~(1998) observed the central radio core in a relatively
quiescent phase finding an intrinsic source size of $\sim2$ mas (major
axis) at 15 GHz and fluxes around 40 mJy (flat spectrum). For these
parameters the jet model gives a jet power of $10^{39.1}$ erg/sec and
a $\gamma_{\rm e}$ of $\sim400$. In addition, the predicted scaling of
the core size ($\propto\nu^{-0.9}$) is consistent with the observed
one (roughly $\propto\nu^{-1}$, taking out the scatter broadening).
The observed time delay of $\sim4$ min of outburst peaks between 3.5
and 2cm can be explained as the delay in time it takes for each
outburst to reach the optically thin regime ($\tau\simeq1$) at the
angular distance $\Phi_{\rm s}$ where the outburst first becomes
visible. For the parameters given here and in Table 1 we
get\footnote{This is smaller than $\Phi$ where the spectrum should
peak. In the light curve the actual peak of an outburst could appear
somewhere between those scales if the duration of the outburst were
smaller than the travel time between $\Phi_{\rm s}$ and $\Phi$.}
$\Phi_{\rm s}=0.035$ mas and the time delay between 2 and 3.5 cm is
predicted by the model to be of order 3 mins. The velocity of the jet
in the model grows asymptotically as determined by Eq.~\ref{euler2}
(see also Falcke 1996b), yielding $\beta=$0.92 at $10^4 R_{\rm g}$ and
$\beta=0.96$ at the scale of a few mas, where the radio emission is
coming from. Considering that Eq.~\ref{euler2} is a no-fit
asymptotical description of the velocity field in the jet this is a
reasonably good prediction.  Clearly, the pressure gradient effect
must be at work at least to some degree here. Mirabel \& Rodriguez
(1994) found tentatively that---in addition to their advance
speed---the blobs may also expand with $0.2c$ at larger scales, thus
perhaps finding direct evidence for a relativistic ``sound speed''
which is needed for the pressure gradient effect to be important.  All
in all the Falcke (1996b) model for M81* seems to give a remarkable
good description of GRS~1915+105 as well and it confirms the basic
Hjellming \& Johnston (1988) picture for compact radio cores in
stellar mass black hole candidates.

\subsection{M81* and Sgr A*}
Finally, for a consistency check, we will apply the model presented
here also to M81* and Sgr A* for which we have discussed very similar
jet-models and their parameters already in earlier papers (Falcke,
Mannheim, Biermann 1993 and Falcke 1996b) while adding a few recent
results that have appeared in the literature.

For M81* Ebbers et al.~(1998) and Bietenholz et al.~(1998) presented
some new measurements confirming that indeed it most likely has a
core-jet structure. The average flux and size of M81* at 8.5 GHz was
$\sim120$ mJy and $\sim0.5$ mas. Falcke (1996b) concluded that a range
of inclinations between $30-40^\circ$ fitted the radio observations
best and this was confirmed by the detection of a nuclear emission
line disk in M81 with similar inclination (Devereux et al.~1997). The
jet power derived from our model of M81* with these parameters is
$10^{41.8}$ erg/sec and $\gamma_{\rm e}=250$.  For comparison, Ho et
al.~(1996) give a bolometric nuclear luminosity for M81 of $10^{41.5}$
erg/sec.

For the Galactic Center radio core Sgr A* observational results now
convincingly demonstrate the presence of a black hole of mass
$2.5\cdot10^6M_{\sun}$ (Ghez et al.~1998; Eckart \& Genzel 1996) while
the exact nature of this source remains ambiguous. For the intrinsic
size of Sgr A* at 43 GHz Bower \& Backer (1998) give a 2$\sigma$ value
of $\sim 0.5$mas and in a later paper Lo et al.~(1998) indeed claim
this to be the intrinsic size together with an elongated source
structure they find. Assuming an arbitrary inclination angle of
45$^\circ$ one can fit Sgr A* with a jet power of $10^{38.7}$ erg/sec
and $\gamma_{\rm e}=125$. The predicted spectral index of
$\alpha=0.17$ is also consistent with observed values (Falcke et
al.~1988) and with the electron energies we found one could in
principle explain the (sub)mm-bump in the spectrum as emission arising
from the inner nozzle region of the jet (see Falcke 1996a). On the
other hand there is currently no evidence for any emission of Sgr A*
at other wavelengths than the radio, suggesting that any accretion
``disk'' emission is well below $10^{38}$ erg/sec and
therefore---unlike in the other sources---is well below the required
jet-power.

\begin{table*}
\begin{center}
\begin{tabular}{l|cccccc|c|ccc}
Source & $D$ & $i$ &$\nu_{\rm obs}$ & $S_\nu$ & size & $M_\bullet$
&$L_{\rm disk}$ & $Q_{\rm jet}$ & $\lg \gamma_{\rm e}$ & $\alpha$\\
       &     &  &[GHz]              & [mJy]   & [mas]& [$M_\odot$]& [erg/sec]      & [erg/sec]     &            &      \\
\hline
GRS~1915+105*&12 kpc& 70$^\circ$& 15 & 40& 2 & (33) & $10^{39}$ &
$10^{39.1}$ & 2.6 & 0.21\\
NGC~4258*&7.3 Mpc& 82$^\circ$& 22 & 3 & 0.35& $3.5\cdot10^7$& $(10^{42})$ & $10^{41.7}$ & 2.8 &0.22\\
M81*     &3.25 Mpc& 35$^\circ$& 8.5& 100 & 0.5& $(10^6)$ & $10^{41.5}$
& $10^{41.8}$ & 2.4 & 0.14\\
Sgr A*  &8.5 kpc& (45$^\circ$) & 43& 1100 & $\sim$0.5 & $2.5\cdot10^6$&$<10^{38}$ & $10^{38.7}$ & 2.1&0.17\\
\end{tabular}
\end{center}
\caption[]{Parameters for compact radio core in various
sources. Columns 2-7 are observationally determined input parameters:
distance $D$, inclination angle $i$ of disk axis and jet to the line
of sight, observing frequency $\nu_{\rm obs}$, flux density of radio
core $S_\nu$, size of radio core, and black hole mass $M_\bullet$. The
inferred disk luminosity $L_{\rm disk}$ is not an input parameter here
and given in column 8 for comparison only. Uncertain values are given
in brackets, but since the black hole masses do not enter strongly the
uncertainties in the black hole mass for M81 and GRS~1915 are actually
irrelevant. Columns~9-11 are output parameters of the radio core
model: jet power $Q_{\rm jet}$, characteristic electron Lorentz factor
$\gamma_{\rm e}$, and average spectral index $\alpha$
($S_\nu\propto\nu^\alpha$) in the radio.}
\end{table*}

\section{Discussion and Conclusion}
\subsection{Model fitting}
The results of the model fitting in the previous section has a number
of interesting consequences and caveats. The fact that all radio cores
can be fitted with one simple model is already important, since the
sources discussed here are so well constrained that by far not all
combinations of size and flux could be fitted by the model. What is,
however, more striking is the fact that the parameters required to
explain sizes and fluxes are very similar. Firstly, in all sources the
typical Lorentz factor of the electrons is of the order of a few
hundred. Given the extreme simplicity of the model and the extensive
use of equipartition assumptions (departing from which would be
reflected in a change of $\gamma_{\rm e}$ as well) this similarity
points to a relatively similar internal structure of the radio cores.
Secondly, all sources have jet-powers very close to or larger than the
luminosity of their thermal radiation, i.e.~the suspected accretion
disk luminosity. Since the model was constructed such that the
radiative efficiency is maximal, applying more realistic models---for
example by using different equipartition factors, velocity fields, or
electron distributions---will therefore in almost all cases only lead
to an increased demand for jet power in these sources.

The model fits perhaps most convincingly the jet in GRS~1915+105,
where it not only reproduces core size and flux very well, but
apparently also predicts radio time delay and jet velocity reasonably
well. The latter indicates that perhaps the pressure gradient in the
jet of GRS~1915+105 is mainly responsible for reaching its
asymptotical velocity---unless higher velocities are found in future
observations.

\subsection{Limitations}
We note that for determining $Q_{\rm jet}$ from the jet model, the
flux and to some degree the inclination angle (especially for small
$i$) are most important, while $\gamma_{\rm e}$ is mainly determined
by the size of the core. Consequently, the latter seems to be the most
uncertain part since it is often ambiguous how to define the core and
its characteristic size, especially when the resolution is of the
order of the core size. Moreover, we have introduced $\gamma_{\rm e}$
mainly to easily reflect the possibility of a low-energy cut-off or
break in the spectrum. This has the positive effect that cores can be
larger than their size purely given by the $\tau=1$ surface (which is
particularly useful in GRS~1915+105 and NGC~4258), however, it also
means that the core size may depend sensitively on the evolution of
the electron distribution which we have ignored almost
completely. Hence, the predictive power of this model for radio core
sizes is very limited and only good to an order of magnitude, so that
we will not base our interpretation heavily on the sizes. It should
also be noted that the jet model used here has been trimmed towards
LLAGN to achieve the greatest possible degree of simplification with
the assumption that their velocity field can be described by
Eq.~\ref{euler2}. We know that this does not apply to quasars where
the bulk Lorentz factor of the jets seems to be larger and the fully
parametrized equations (e.g. as in Falcke \& Biermann 1995) have to be
used.

However, taking all this into consideration we can give yet a more
simplified formula where we have fixed $\gamma_{\rm e}$ at an
intermediate value of 300 and which can be used to very roughly
estimate the jet power of a LLAGN from its flux and
presumed inclination angle alone:

\begin{equation}
Q_{\rm jet}=10^{39}\,\mbox{erg/sec}\;{\frac{{{0.024}^
       {{\frac{{{\xi }_0}}{{{\xi }_1}}}}}\,
     {{1.56}^
       {{\frac{{{\xi }_4}}{{{\xi }_1}}}}}\,
     {{{\left({D\over10{\rm kpc}}\right)}}^
       {{\frac{1.6}{{{\xi }_1}}}}}\,
     {{{\left({S_\nu\over{\rm mJy}}\right)}}^
       {{\frac{0.79}{{{\xi }_1}}}}}}{
     {{{\left({M_\bullet\over33 M_\odot}\right)}}^
       {{\frac{0.15\,{{\xi }_2}}
          {{{\xi }_1}}}}}\,
     {{{\left({\nu\over8.5 {\rm GHz}}\right)}}^
       {{\frac{0.15\,{{\xi }_2}}
          {{{\xi }_1}}}}}\,
     {{{{\xi }_3}}^
       {{\frac{0.79}{{{\xi }_1}}}}}}}
\end{equation}

\subsection{Jet/disk-symbiosis}
The main result of this work, however, is that in three very different
sources, with very different sizes and fluxes, we can explain the
central core with a single model by just scaling the jet power with
the accretion rate. This works because the three selected sources,
GRS~1915+105, NGC~4258, and M81*, all have some very important
ingredients in common. All three have clear evidence for a massive
black hole, signs of (large or small scale) accretion disks, jet
structures in their radio cores, and a good determination of the
inclination angle (important for the fitting of individual
sources). In GRS~1915+105 there is even direct evidence for a coupling
between jet and disk from the light curves. The high jet power we
derive for the radio core in a relatively quiescent phase is quite
consistent with but lower than the power derived for the major
outbursts (e.g.~Mirabel et al.~1998).

In hindsight this high power in the radio cores justifies the
assumptions we have made in earlier papers that jet and disk can be
considered symbiotic systems and that---at least in a few
systems---the assumption of $Q_{\rm jet}/L_{\rm disk}\sim1$ (or even
larger) seems appropriate. This also strengthens the picture that the
jets are produced in the inner region of an accretion disk, where a
major fraction of the dissipated energy is channeled into the jet
(Falcke \& Biermann 1995; Donea \& Biermann 1996). As a consequence,
modelling of accretion disks and X-ray light curves in jet systems
like GRS~1915+105 clearly requires taking the jet into account.

\subsection{Accretion disks and ADAFs}
Another consequence from those radio cores is their amazing scale
invariance.  It seems that we can use the very same model for a
stellar mass black hole which is accreting near its Eddington limit
(GRS~1915+105) as well as for a super-massive black hole which is
presumably accreting at an extreme sub-Eddington rate (M81,
NGC~4258). Moreover a very similar model was successfully applied to a
quasar sample earlier (Falcke et al.~1995), i.e. super-massive black
holes near the Eddington limit. That would suggest that certain
properties of an accretion disk/flow, namely jet production, is very
insensitive towards changes in accretion rate or black hole mass, and
that the 'common engine' mechanism of black hole accretion and jet
formation, suggested by Rawlings \& Saunders (1991), may include a
much larger range of AGN than only quasars and radio galaxies.

If this is so, it has to be asked whether indeed every accretion flow
necessarily has to make a transition from a thin $\alpha$-disk to an
ADAF when turning sub-Eddington. In order to maintain this proposition
one then needs to explain how the accretion disk structure can change
so drastically without affecting its innermost region, where jets
presumably are being produced. If one asks what the arguments for
ADAFs in low-power AGN really are, the evidence remains thin. For
NGC~4258 it is quite obvious now that the radio core cannot serve to
support an ADAF emission model. Quite contrary the derived jet power
is consistent with a low accretion rate, which in turn is consistent
with the low luminosity of the nucleus and, of course, a thin disk is
directly seen at least at larger radii. With the radio emission gone
the ADAF spectral energy distribution, extending over many orders of
magnitude, merely serves to explain a single X-ray data point. Hence,
it is currently not obvious that an ADAF is really needed to explain
this source at all.

The same seems to be true for M81, which is equally sub-Eddington as
NGC~4258. Here the situation is even worse, since Ishisaki et
al~(1996) also claim the detection of a broad iron Fe-K line
suggesting that probably the inner disk cannot be as hot as required
in an ADAF model. As similar broad line has been tentatively claimed
also for NGC~4258 by Cannizzo et al.~(1997). In any case the two
galaxies serve as a general warning to view the existence of a compact
radio core in low-luminosity AGN as prima facie evidence for an
ADAF---a jet origin may be a more natural explanation here and in
other cases. We also want to point out that the argument by Narayan et
al.~(1995) a pure ADAF interpretation of radio cores were superior
because the latter requires an 'additional' emission component is not
quite true if one considers disks and jets to be symbiotic, i.e. to be
essentially one system.

\subsection{Sgr A*}
So far we have concentrated the discussion mainly on the three sources
which fitted the model well and have not mentioned Sgr A*. The
jet-power we derive for the latter source now is virtually unchanged
with respect to the power used to introduce the Sgr A* jet model in
Falcke et al.~(1993) and it still explains the radio properties of Sgr
A* very well. However, as pointed out in this earlier paper already,
the required jet power also provides a {\it lower limit} to the
accretion rate onto the black hole of $10^{-7} M_\odot/$yr (using the
current numbers and ignoring unlikely small inclination angles). Yet,
even such a low accretion rate seems to be excluded on the basis of
very stringent upper limits for the NIR flux of Sgr A* placed by
Menten et al.~(1996, see also Falcke \& Melia 1997 for a discussion of
this point). If one ignores the possibility of intrinsic obscuration
in the Galactic Center this indeed seems to indicate a very low
radiative efficiency of the accretion flow onto the black hole and,
ironically, the jet model may provide in this case supporting argument
for advection.  As an alternative model one could envisage a scenario
where the missing energy, instead of being radiated away, is put
almost completely into a (magnetically) driven wind throughout the
disk---this, however, would need to be worked out in more detail.  As
a side note we also want to mention that the low $\gamma_{\rm e}$ we
derive for Sgr A* requires a relativistic electron fraction of
$\sim3$.  A value for $x_{\rm e}$ significantly larger than unity is
only possible if additional pairs were produced and thus could support
the suggestions by Falcke (1996b) and Mahadevan (1998) that pair
production through proton-proton collisions is at work here. However,
as pointed out before, the determination of $\gamma_{\rm e}$ and
especially its interpretation within our simple model is fairly
unreliable and hence should be taken very cautiously.

\subsection{Summary}
To summarize this section, we believe that the jet model does seem to
provide an excellent description of nuclear radio cores also in LLAGN
and that considering jets and disks as symbiotic systems can explain
the vast range of radio core luminosities and sizes we find in the
nuclei of galaxies and in stellar mass black holes. Comparison of jet
powers and nuclear luminosities of some radio cores seem to indicate
that they are of similar order, thus supporting an under-fed black hole
scenario with low accretion rates. Consequently, even though they are
not excluded, ADAFs need not be as ubiquitous in low-luminosity AGN as
has been claimed while the jet interpretation for compact radio nuclei
seems to be a natural interpretation also for low-luminosity AGN.  It
therefore does not appear as if the presence of a compact radio core
and a low optical luminosity alone serve as a good indicator for an
ADAF. On the other hand---as the example of Sgr A* shows---a jet
interpretation of low-luminosity radio cores could in some cases
support the presence of an advection dominated accretion flow or some
other kind of radiatively deficient accretion disk. Therefore a
combination of ADAF and jet models should also be considered for the
fitting of nuclear spectral energy distributions.

\acknowledgements 
HF is supported by DFG grant Fa 358/1-1\&2.

\end{document}